%% file: sn-article.tex
\documentclass[iicol]{sn-jnl}

\usepackage{graphicx}%
\usepackage{multirow}%
\usepackage{amsmath,amssymb,amsfonts}%
\usepackage{amsthm}%
\usepackage{mathrsfs}%
\usepackage[title]{appendix}%
\usepackage{xcolor}%
\usepackage{textcomp}%
\usepackage{manyfoot}%
\usepackage{booktabs}%
\usepackage{algorithm}%
\usepackage{algorithmicx}%
\usepackage{algpseudocode}%
\usepackage{listings}%
\usepackage{framed}

\theoremstyle{thmstyleone}%
\theoremstyle{thmstyletwo}%

\theoremstyle{thmstylethree}%

\begin{document}

\title[Article Title]{Large Language Models in the Data Science Lifecycle: \\A Systematic Mapping Study}



\author[1]{\fnm{Sai Sanjna} \sur{Chintakunta}}\email{sqc6557@psu.edu}

\author[1]{\fnm{Nathalia} \sur{Nascimento}}\email{nqm5742@psu.edu}
\equalcont{These authors contributed equally to this work.}

\author[1]{\fnm{Everton} \sur{Guimaraes}}\email{ezt157@psu.edu}
\equalcont{These authors contributed equally to this work.}

\affil[1]{\orgdiv{Engineering Division, Great Valley}, \orgname{The Pennsylvania State University}, \orgaddress{\state{Pennsylvania}, \country{United States}}}


\abstract{In recent years, Large Language Models (LLMs) have emerged as transformative tools across numerous domains, impacting how professionals approach complex analytical tasks. This systematic mapping study comprehensively examines the application of LLMs throughout the Data Science lifecycle. By analyzing relevant papers from Scopus and IEEE databases, we identify and categorize the types of LLMs being applied, the specific stages and tasks of the data science process they address, and the methodological approaches used for their evaluation. Our analysis includes a detailed examination of evaluation metrics employed across studies and systematically documents both positive contributions and limitations of LLMs when applied to data science workflows. This mapping provides researchers and practitioners with a structured understanding of the current landscape, highlighting trends, gaps, and opportunities for future research in this rapidly evolving intersection of LLMs and data science. }

\keywords{Large Language Models, Data Science, Systematic Mapping Study}



\maketitle

\input{sections/1-introduction}
\input{sections/2-background}
\input{sections/3-methodology}

\input{sections/4-results}
\input{sections/5-implications}
\input{sections/6-limitations-and-treats}

\input{sections/7-conclusion}

\bibliographystyle{splncs04}
\bibliography{daan-bibliography}

\end{document}

%% file: sections/1-introduction.tex
\section{Introduction}
Large Language Models (LLMs) have been applied to a broad range of problems, demonstrating capabilities that go beyond merely interpreting natural language instructions \cite{nascimento2025llm}. They have shown the ability to support dynamic problem-solving, adapt to diverse contexts, and integrate with computational tools to perform complex, multi-step tasks \cite{halevy2023will}. Their applications range from code generation in software engineering \cite{hou2024software} to decision-making in the clinical domain \cite{foltyn2024gpt4}. This versatility has sparked growing interest in their potential within Data Science, where tasks often go beyond simple question answering or code synthesis \cite{gptdatascience} and typically require both domain expertise and programming skills.

A data science lifecycle \cite{instituteData2025lifecycle,datasciencelifecycle} structures these activities into interconnected stages that guide work from problem formulation through evaluation and into production, providing a process to extract value from data. In this paper, we adopt the five-stage model presented in \cite{instituteData2025lifecycle} as our reference, which organizes the process into Problem Definition, Data Collection and Preparation, Data Exploration and Analysis, Model Building and Evaluation, and Deployment and Maintenance. This framework captures the diversity and technical demands of data science while allowing adaptation to specific project requirements.

Although LLMs have been applied to specific tasks within the data science lifecycle, evaluations have typically occurred in isolated contexts. Despite this growing interest, there remains a gap in the literature for comprehensive surveys that holistically examine the impact of these models across the entire lifecycle. Understanding the role of LLMs across the data science lifecycle is critical as these models increasingly shape how data is accessed, transformed, and interpreted. Their integration into tasks such as database querying, data preprocessing, and analytical reasoning \cite{halevy2023will, gptdatascience} signals a shift toward workflows where natural language can drive complex operations. Investigating their impact holistically can reveal both the opportunities for democratizing advanced analytics and the challenges in ensuring reliability, transparency, and domain-specific applicability.

To comprehensively examine the landscape of LLM applications across the data science lifecycle, we conducted a systematic mapping study (SMS) \cite{wohlin2012experimentation}. An SMS is suited to broad questions and less explored fields, which is appropriate given the emergence of LLMs in data science. It surveys a wide body of work to provide a categorized overview of the state of the art. Following predefined protocols, we searched, selected, and classified relevant studies to build a map of the literature, emphasizing the classification and quantification of research outputs to identify clusters, trends, and gaps. To the best of our knowledge, this is the first SMS analyzing the applications and impact of Large Language Models across all five stages of the data science lifecycle. The results lay the groundwork for targeted future studies and provide practitioners with insights into how LLMs can most effectively transform data science workflows and outcomes. The contributions of this paper are as follows:

\begin{itemize}
    \item A comprehensive systematic mapping study analyzing LLM applications across all five stages of the data science lifecycle.
    \item A synthesis of assessment methods, datasets, and performance metrics used to evaluate LLMs.
    \item An overview of the LLM families studied and their application domains.
    \item An impact assessment of positive and negative effects on data science processes and outcomes.
    \item Identification of research gaps, current limitations, validity threats, and future research directions.
\end{itemize}

%% file: sections/2-background.tex
\section{Background}

\subsection{Data Science Lifecycle}
A data science lifecycle is a structured process to extract value from data, guiding work from problem formulation through evaluation and into production. Different workflows and frameworks have been proposed to describe the data science lifecycle, including KDD, CRISP-DM, and Microsoft’s nine-stage workflow \cite{KDD, CRISP, Amershi}. They share commonalities centered on data analysis and processing tasks. Although there is no universally accepted lifecycle model for data science projects \cite{biswas2022art,datasciencelifecycle,lin2018design}, these approaches converge on recurring stages that combine data-oriented activities (e.g., data collection, cleaning, labeling) and model-oriented activities (e.g., feature engineering, training, evaluation, deployment, monitoring). In this paper, we adopt the five-stage model presented in \cite{instituteData2025lifecycle}: Problem Definition, Data Collection and Preparation, Data Exploration and Analysis, Model Building and Evaluation, and Deployment and Maintenance. This model captures the main stages and allows adaptation to project requirements and complexity. 

In line with \cite{instituteData2025lifecycle}, the main activities typically addressed in each stage are: (i) Problem Definition: clarify objectives, requirements, constraints, and success criteria, often via stakeholder interviews and problem framing; (ii) Data Collection and Preparation: identify and acquire data from relevant sources, address privacy and governance, and prepare the data through cleaning, integration, feature engineering, and dimensionality reduction; (iii) Data Exploration and Analysis: conduct exploratory analysis and visualization, compute descriptive statistics, and apply initial modeling to validate assumptions, uncover patterns, and detect anomalies; (iv) Model Building and Evaluation: select suitable algorithms, train models, and evaluate performance using appropriate metrics (e.g., accuracy, precision, recall, F1), while assessing bias and interpretability; and (v) Deployment and Maintenance: plan deployment (e.g., batch, real time, or cloud), monitor performance, detect data drift, update models, and maintain security controls. 

For the search string used in this systematic mapping study (Section \ref{sec:research}), we addressed variations in nomenclature, and we included tasks commonly associated with each stage to cover alternative representations in the literature.

\subsection{Large Language Models (LLMs)}
LLMs are Transformer-based neural networks trained on large text corpora to perform a wide range of natural language processing tasks. Depending on their architecture, LLMs may follow an encoder–decoder, decoder-only, or encoder-only design, with decoder-only models dominating modern generative applications. While there are many LLMs, some of the most widely used ones are the GPT series by OpenAI \cite{openai2023gpt}, Llama \cite{touvron2023llama}, and Gemini \cite{team2023gemini}.

Beyond text generation, LLMs—especially at larger scales (e.g., exceeding 100 billion parameters)—have demonstrated utility in analytical workflows and decision-making, supporting tasks such as data preprocessing, querying, and interactive analytics across the data science lifecycle \cite{gptdatascience}.

\section{Related Work}
Previous studies have examined the impact of Large Language Models on data science \cite{gptdatascience,abumalloh2024, havely2023-data-science}, revealing their capacity to automate various aspects of the workflow while highlighting both advantages and limitations. 

Literature reviews, such as Hassani et al. \cite{hassani2023chatgpt}, have primarily focused on ChatGPT, with occasional comparisons to models like BERT, detailing the strengths and weaknesses of each. These reviews document LLM impacts on tasks such as data augmentation, text mining, natural language processing, synthetic data generation, and programming. The paper addresses software engineering programming capabilities of LLMs rather than data science, focusing on topics like data structures and algorithms.  \newline While some researchers have attempted to address the entire data science lifecycle, many surveys concentrate narrowly on specific tasks. For instance, Zhang et al. \cite{zhang2024} surveyed natural language interfaces, particularly LLMs, for visualization and tabular data analysis, while Ding et al. \cite{ding2024} examined data augmentation techniques using Large Language Models, categorizing approaches into data creation, labeling, reformation, and co-annotation. Another study \cite{zhen2024} focuses on data annotation, a crucial task in natural language processing. The survey provides an overview of the effects of LLMs on data annotation. 
 Beyond literature reviews and surveys, numerous papers have evaluated LLMs through experimental assessments across diverse application domains. Arasteh et al. \cite{xie2024waitgpt} demonstrate the use of ChatGPT Advanced Data Analysis to build and implement machine learning models, including Gradient Boosting Machine and logistic regression, for clinical tasks. Additionally, many studies introduce specialized frameworks incorporating LLMs for specific data science tasks (e.g.,  DSBot  \cite{chen2025viseval}, Chat2VIS \cite{maddigan2023chat2vis}, HuggingGPT \cite{arasteh2024langchain}). 

Despite these valuable contributions, existing research exhibits notable limitations. Studies like \cite{hassani2023chatgpt} focus on specific models such as ChatGPT, constraining the breadth of the examined LLMs. Other survey papers \cite{ding2024, zhen2024, zhang2024} address only selected stages of the data science lifecycle rather than adopting a comprehensive approach. Moreover, the literature remains fragmented, lacking a unified perspective on how different LLM architectures impact the full spectrum of data science processes.
Our study addresses these gaps through a systematic mapping study that categorizes previous work according to specific stages in the data science lifecycle while comprehensively analyzing various LLM architectures and their distinct applications. This approach enables a more holistic understanding of how different models impact specific data science processes, providing a foundation for future research and practical implementations.

%% file: sections/3-methodology.tex
\section{Research Methodology} \label{sec:research}
To investigate the impact of LLMs on the data science lifecycle, we formulated our research questions using the Goal-Question-Metric (GQM) paradigm \cite{gqm}. The GQM approach establishes a clear hierarchy between our research objectives, specific questions, and measurable metrics for analysis.
The overarching goal of this study is to systematically examine the impact of various LLMs on the data science lifecycle. From this goal, we derived the following research questions and corresponding metrics.

The methodology began with the development of a comprehensive search string that was executed across IEEE Xplore and Scopus. This initial search yielded 172 papers that required filtering to identify the most relevant studies. 

\begin{figure}[h]  
    \centering
    \includegraphics[width=0.4\textwidth]{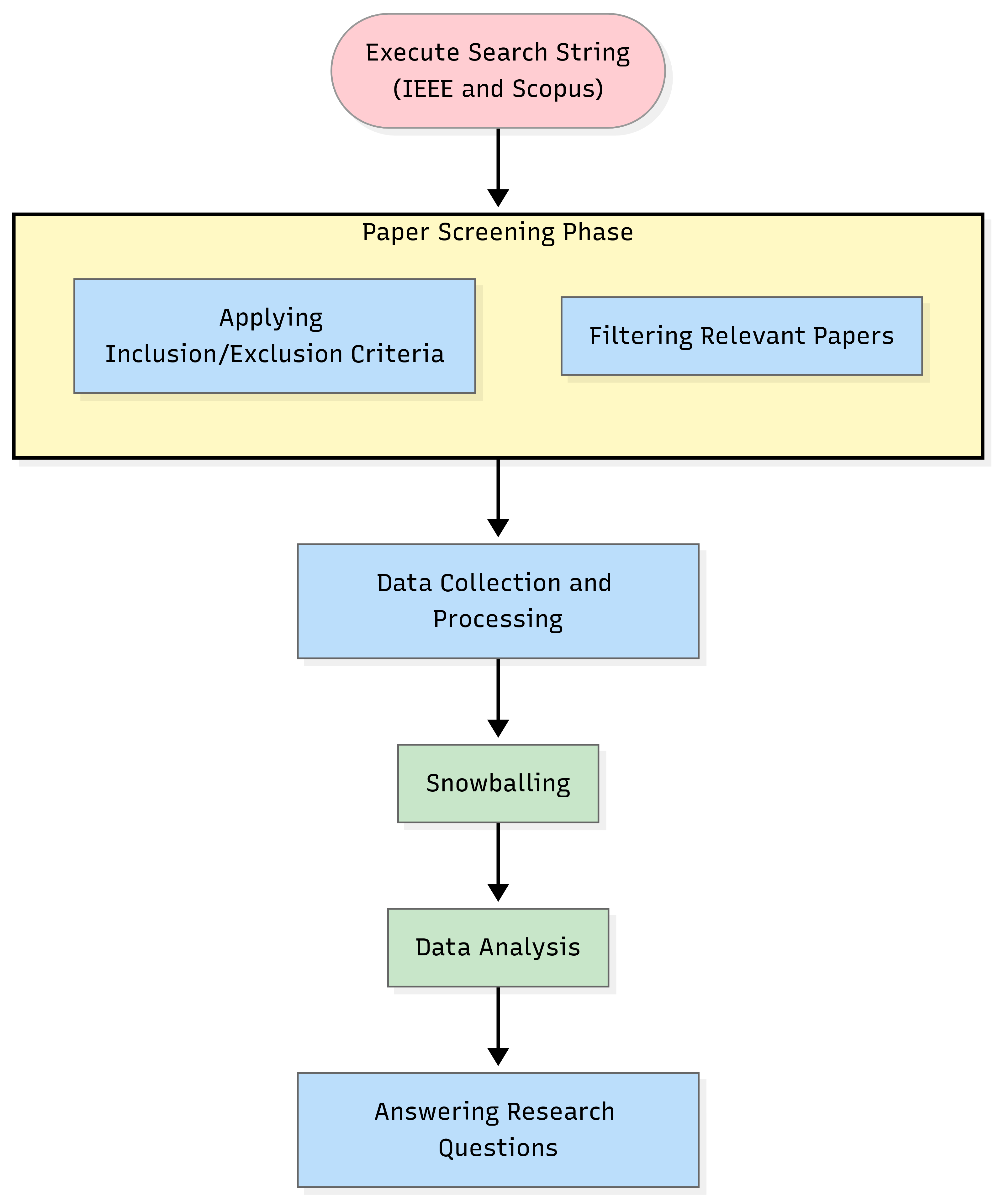}
    \caption{Methodology}
    \label{fig:my_label}
\end{figure}

Following the initial search, we applied the inclusion and exclusion criteria as defined in Section \ref{subsec:criteria} to filter the retrieved papers. The paper screening phase consisted of two key steps: first, applying the predefined inclusion and exclusion criteria to remove irrelevant papers, and second, filtering the remaining papers based on their relevance to the research topic. The number of papers obtained from the initial search and remaining after each filtering step is shown in Figure \ref{fig:my_label}. 

Next, we conducted a detailed review of each paper's abstract and introduction to determine its relevance to our research objectives. Once the final set of papers was obtained through the screening process, we performed snowballing techniques (both forward and backward citation analysis) to identify any additional relevant papers that may have been missed in the initial search. This step helps ensure comprehensive coverage of the literature by following citation networks to discover related work.

Following the completion of the snowballing process and finalization of our paper dataset, we conducted a thorough review of each selected paper. During this data collection and processing phase, we extracted the necessary information and data points required to answer our research questions. 
The final phase involved comprehensive data analysis to synthesize the collected information and provide evidence-based answers to our research questions, ultimately contributing to our understanding of how LLMs impact various stages of the data science lifecycle.

\subsection{Research Questions}
We carefully designed six research questions to systematically review the literature and comprehensively complete this mapping study. Each question addresses a key aspect, ensuring thorough coverage. \newline
\textbf{RQ1. What data science stages and tasks have been addressed by LLMs?} 
\newline The RQ1 seeks to identify and categorize the main stages of the data science lifecycle where LLMs have been applied. This is done by systematically classifying published studies by data science stage (e.g., data collection, preprocessing, analysis, visualization) with frequency analysis of task types within each stage.
\par \textbf{RQ2. In what ways have LLMs been evaluated for data science tasks, and what are the datasets used to assess them?} The RQ2 is drafted to determine the evaluation approaches, datasets, and metrics used to assess LLM performance in data science tasks. This is accomplished through the synthesis of evaluation methodologies, including: (1) binary variable to categorize empirical study, (2) classification of evaluation approaches (e.g., case study, controlled experiment), (3) catalog of datasets, and (4) performance metrics used across studies.
\par \textbf{RQ3. What LLMs are being analyzed for data science tasks?} The RQ3 identifies the specific Large Language Models being studied for data science applications and whether they are fine-tuned. This is answered by keeping track of: (1) Type of Large Language Model (e.g., GPT, Llama), and (2) Binary value to check if fine-tuning is performed.
\par \textbf{RQ4. What are the application domains of LLMs for data science tasks?} The RQ4 seeks to identify the various domains of applications where LLMs are employed for data science tasks. This is done by domain classification identifying: (1) general-purpose applications, (2) domain-specific applications, and (3) cross-domain patterns in application types.
\par \textbf{RQ5. What are the reported positive and negative impacts of LLMs on the data science lifecycle?}
The goal of the RQ5 is to analyze the reported positive and negative impacts of LLMs on data science processes and outcomes as reported in the literature. This is achieved through a thematic analysis of reported impacts, which identifies the frequency of reported benefits and challenge types, as well as the correlation between model characteristics and reported impacts.

\par \textbf{RQ6. What are the reported research gaps when applying LLMs in the data science lifecycle?} 
\newline
The RQ6 seeks to analyze the reported research gaps, validity threats, and future research directions in the application of LLMs to data science, as reported by the authors. This is achieved through the identification and qualitative analysis of research gaps, validity threats, and future research directions.

\subsection{Search String}

To ensure that the research is reproducible, a standardized search string is generated. The search string, executed in Scopus and IEEE Xplore, is shown below.
\begin{framed}
{TITLE-ABS-KEY ( ( ( ``data science" OR ``data analytics" ) AND ( ``Generative AI" OR ``Large Language Models" OR ``LLM" ) AND ( ( ``data visualization" OR ``chart generation" OR ``plot creation" ) OR ( ``code generation" OR ``script generation" OR ``automated coding" ) OR ( ``data preparation" OR ``data cleaning" OR ``data wrangling" OR ``data preprocessing" OR "ETL" ) OR ( ``data analysis" ) OR ( ``Model Deployment" OR ``Model Implementation" OR ``Model Maintenance" OR ``Model Monitoring" OR ``Deployment" ) ) ) )} 
\end{framed}

\begin{figure}[h]  
    \centering
    \includegraphics[width=0.3\textwidth]{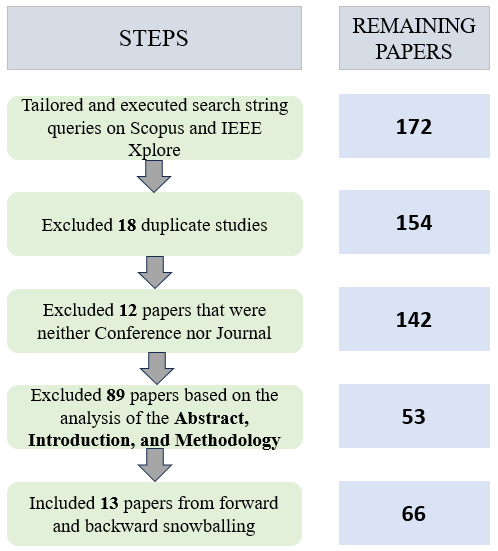}
    \caption{Paper Screening Process}
    \label{fig:searchstring}
\end{figure}
 
The search string yielded 87 documents from Scopus and 85 from IEEE Xplore. Each retrieved paper was screened to assess its relevance to the study. The screening process began with an evaluation of the abstracts, which were classified as relevant, not relevant, or unsure (requiring further inspection), as shown in Figure \ref{fig:searchstring}. The second step involved reviewing the introductions of the papers. Finally, the full texts were read to make a comprehensive assessment.

\subsection{Inclusion and Exclusion Criteria}\label{subsec:criteria}
To ensure methodological rigor and relevance of our systematic mapping study, we established clear inclusion and exclusion criteria before conducting our literature search. These criteria guided our paper selection and review process.

\textbf{Inclusion Criteria}
We included papers that met all of the following criteria:
(i)  published after January 2020, capturing the most recent advancements in LLM applications for data science; (ii) explicitly evaluated the performance, impact, or integration of LLMs within one or more data science lifecycle stages, (iii) published in peer-reviewed conferences, journals, or as preprints in established repositories; and (iv) paper and prompts written in English to ensure comprehensive analysis.

\textbf{Exclusion Criteria}
We excluded papers that met any of the following criteria: (i) published before 2020, as we aimed to focus on current LLM capabilities and applications; (ii) papers that did not contain any substantial evaluation of LLMs' impact on data science tasks or processes; (iii) employed non-English prompts for LLM interactions, which would limit replication and comparison; and (iv) focused primarily on software engineering lifecycle rather than data science workflows.

During our search process, we screened titles and abstracts to exclude papers that studied LLMs in contexts unrelated to data science (e.g., code generation for software development, literary analysis, conversational agents). The final corpus was then subjected to full-text review to ensure alignment with our research objectives.

%% file: sections/4-results.tex
\section{Systematic Mapping Results}
In this section, we present and discuss the results gathered from the systematic mapping study. The supplementary material \cite{chintakunta202516816445} contains the complete list of papers identified in our search, along with their answers to the associated variables used to address the research questions, such as the LLMs employed, the stages of the data science lifecycle addressed, the tasks performed, the application domains, and the evaluation approaches. Each subsection below provides insights and answers to the research questions defined in our study.

\subsection{RQ1: Data Science Lifecycle Stage}
\begin{figure}[h]  
    \centering
    \includegraphics[width=0.4\textwidth]{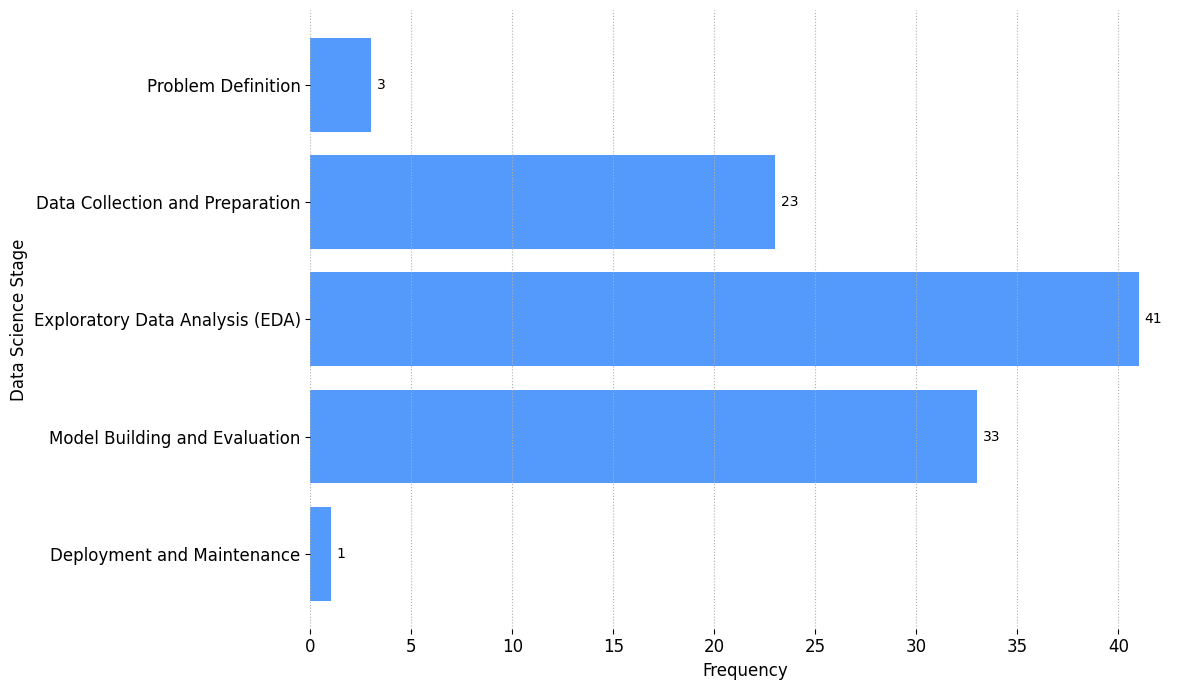}
    \caption{Frequency of Data Science Stages Addressed}
    \label{fig:rq1_lifecycle}
\end{figure}

\begin{figure}[h]  
    \centering
    \includegraphics[width=0.4\textwidth]{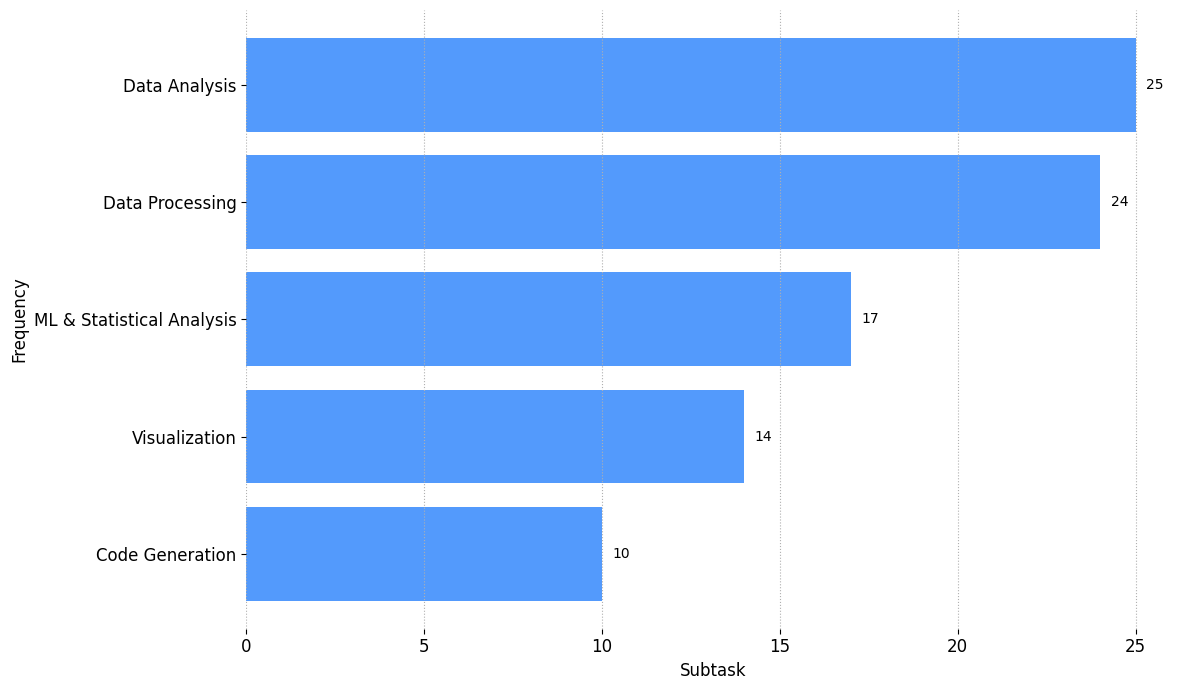}
    \caption{Frequency of Sub-tasks Addressed}
    \label{fig:rq1_sub_task}
\end{figure}

Figure~\ref{fig:rq1_lifecycle} depicts the frequency of the data science stages covered by the papers. When interpreting the graph, it is essential to consider that each paper can address multiple stages of the data science lifecycle. Therefore, the total number represented in the graph does not reflect the complete count of all reviewed papers. The graph indicates that the majority of the papers focus on evaluating LLMs in data analysis tasks (41), followed by model building and evaluation (33), as well as data collection and preparation (23). However, model deployment (1) and problem definition (3) remain underexplored areas within the data science lifecycle. 
Under each stage, we identified several sub-tasks in the reviewed papers, such as data filtering, transformation, sorting, goal understanding, feature engineering, insights generation, and image classification. Figure \ref{fig:rq1_sub_task} shows the most common groups of sub-tasks found in these studies.

The following sections detail the tasks evaluated under each stage.

\subsubsection{Problem Definition}

The initial phase of any data science project involves understanding the problem domain and establishing clear business requirements. This foundational step encompasses three components: (1) identifying and precisely defining the problem requiring resolution, (2) evaluating the feasibility of data science methodologies for addressing the identified challenge, and (3) determining appropriate data sources while establishing a systematic framework for project execution.
This phase has been examined across multiple studies 
\cite{bednar2023cognitive, patel2024,zhong2024}, with particular attention to the comparative capabilities of large language models versus human expertise. Bednár et al. \cite{bednar2023cognitive} conducted a comprehensive analysis comparing GPT-4's performance against human data scientists in problem identification and key performance indicator (KPI) recognition within the aluminum production domain. Their findings revealed complementary competency patterns: while human data scientists demonstrated superior performance in identifying domain-specific technical KPIs, GPT-4 exhibited enhanced capabilities in task definition and problem articulation, suggesting potential for human-AI collaboration in this phase.

\subsubsection{Data Collection and Preparation}

The data collection and preparation phase involves the systematic acquisition, cleaning, and preprocessing of data to ensure high-quality inputs for subsequent analysis. Data wrangling, a critical component of this stage, has been extensively studied in recent literature \cite{huang2024, jaimovitch2023can}.

Huang et al. \cite{huang2024} introduced CoCoMine, a comprehensive framework designed to automate the data wrangling process. CoCoMine achieves precise data transformations by incorporating multiple contextual layers, including code contexts, textual contexts, and data contexts (encompassing both input and output specifications). This multi-contextual approach demonstrates significant improvements in automation accuracy compared to traditional methods.

Nasseri et al. \cite{nasseri2023LLM} evaluated GPT-3.5 Turbo's performance on practical data preparation tasks, including product name translation and categorical assignment using business datasets. Their results indicated that ChatGPT achieves high accuracy in these tasks and demonstrates efficiency across various data formats (text, tabular data). However, consistent with broader research findings, the model struggled with tasks requiring specialized domain knowledge.
Similarly, Ghazzai et al. \cite{ghazzai2024harnessing} assessed GPT-4's performance on the Transform Data by Example (TDE) benchmark, which encompasses diverse data transformation tasks such as temperature conversion (Fahrenheit to Celsius), timezone conversion (EST to PST), and URL parsing for domain extraction. GPT-4 demonstrated superior performance compared to existing transformation systems, highlighting its potential for automated data preprocessing.

\subsubsection{Exploratory Data Analysis}

Exploratory Data Analysis (EDA) represents the most extensively studied phase in the literature, with 41 papers addressing various aspects of this critical stage. Research approaches have varied significantly, addressing different challenges and methodological perspectives.

The predominant research focus has been on automating the data analysis process to enhance accessibility for domain experts \cite{huang2024chatgpt, pons2025,sudha2024}, recognizing that traditional data analytics methods often require specialized programming expertise in languages such as Python and R. Some studies focused on evaluating LLMs through the CRISP-DM framework \cite{musazade2024exploring, patel2024}. For instance, Patel et al. \cite{patel2024} systematically evaluated ChatGPT-4's performance in conducting analysis tasks following the CRISP-DM framework, including the generation of descriptive statistics (standard deviations, correlation matrices, covariance matrices). Chugh et. al. \cite{chuch2023, hu2024infiagent} employ and evaluate LLMs as agents to perform data analysis. 

A substantial subset of research has concentrated specifically on tabular data analysis 
\cite{dong2024on, fang2024llms, liu2023jarvix, shinde2024, sudha2024}. Dong et al. \cite{dong2024on} conducted a comprehensive evaluation of various tabular data tasks, including table matching, cleaning, augmentation, search, and transformation. Additionally, researchers \cite{sudha2024} explored the application of Phind-CodeLlama, a specialized code generation model based on Code Llama, for generating insights from CSV datasets. \cite{dong2024on, liu2023jarvix, ojha2024} focus on text-to-SQL conversion tasks.

Despite these advances, empirical findings consistently indicate that while LLMs demonstrate competency in performing simple tasks and preliminary analysis, human experts maintain superior performance in conducting nuanced and in-depth analytical work \cite{fang2024llms, jiang2024}. This suggests that current LLM applications are most effective as supportive tools rather than complete replacements for human expertise.

\subsubsection{Model Building and Evaluation}

The model-building and evaluation phase has attracted significant research attention, with studies focusing on specific methodological approaches, including linear and logistic regression, feature engineering, and automated code generation for machine learning model development \cite{ahn2024data, foltyn2024gpt4, gramacki2024, khadka2024}. Studies also focused on LLMs as data science programming assistants \cite{Sheese_2024}.

Multiple studies have evaluated LLMs' capability to generate Python code for constructing various predictive models. Khadka et al. \cite{khadka2024} systematically compared different prompting strategies for GPT-4 code generation, including one-shot prompting, step-wise sequential prompting, and bi-sequential prompting, applied to linear regression and random forest models within the Building Management System domain. Their comparative analysis provided insights into optimal prompting methodologies for different modeling contexts.
Foltyn-Dumitru et al. \cite{foltyn2024gpt4} evaluated ChatGPT's Advanced Data Analytics package for building machine learning models using radiomics datasets to predict glioma molecular subtypes. This application demonstrated the potential for LLMs in specialized medical domains while highlighting the importance of domain-specific validation.

Beyond the development of statistical and machine learning models, a significant research focus has been on evaluating LLMs' code generation capabilities for data visualization \cite{gao2025fine, lai2023ds1000, maddigan2023chat2vis, nissen2024utilizing, valverde-rebaza2024, zhao2025leva, zhao2025lightva}. Visualization generation represents a crucial component of the data science lifecycle, serving multiple purposes from initial data exploration to final result presentation. Several studies have proposed specialized frameworks, such as Chat2Vis \cite{maddigan2023chat2vis}, which leverages LLMs to automatically generate visualizations based on structured prompts, demonstrating the potential for automated visual analytics workflows.

\subsubsection{Deployment and Maintenance}
The deployment of machine learning models is a relatively underexplored topic, with only one paper addressing it. Zhong et. al. \cite{zhong2024} conduct a user study with business students to help them solve logistic regression in R. They cover all the phases of the lifecycle, including model deployment.

\subsection{RQ2: Evaluation techniques and Datasets}

\begin{figure}[h]  
    \centering
    \includegraphics[width=0.48\textwidth]{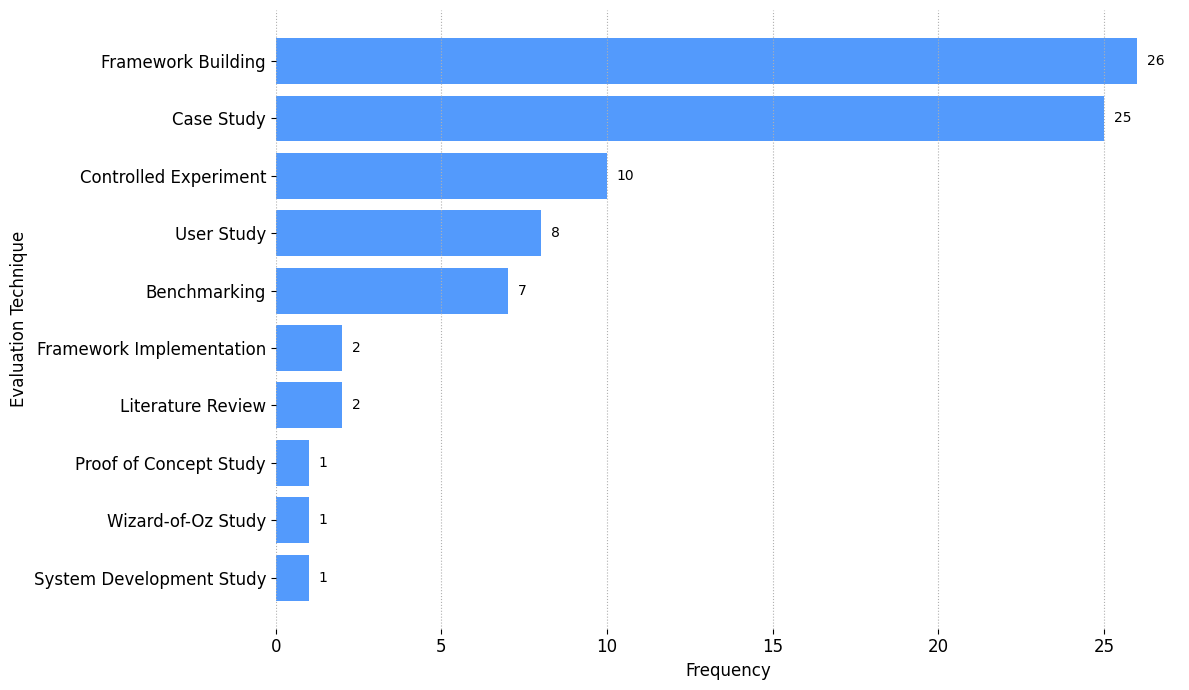}
    \caption{Frequency of Evaluation Techniques}
    \label{fig:rq2_eval_technique}
\end{figure}

Figure~\ref{fig:rq2_eval_technique} reveals a consistent pattern wherein researchers introduce novel frameworks \cite{yuan2025pestgpt} while employing rigorous evaluation methodologies to validate their effectiveness. These frameworks typically combine innovation with systematic assessment. 87.7\% of the reviewed papers conducted an empirical study.
Multiple evaluation techniques are commonly employed within individual studies. Notably, Guo et al. \cite{guo2024dsagentdatascience} introduced DSAgent, a GPT-powered system designed to address complex data science challenges, including regression and classification tasks, thereby enabling the construction of autonomous machine learning models. Similarly, Beasley et al. \cite{beasley2024pipeline} developed LLM4V, providing an end-to-end solution for comprehensive data analysis and visualization.
The majority of these frameworks incorporate robust evaluation protocols. For instance, Zhao et al. \cite{zhao2024new} presented ARGUMENT2CODE, an innovative framework for automating qualitative coding of textual data. To assess this framework, the authors conducted a controlled experiment utilizing the ArgKP-2021 dataset with a 70/30 training-testing split.

\begin{figure}[h]  
    \centering
    \includegraphics[width=0.48\textwidth]{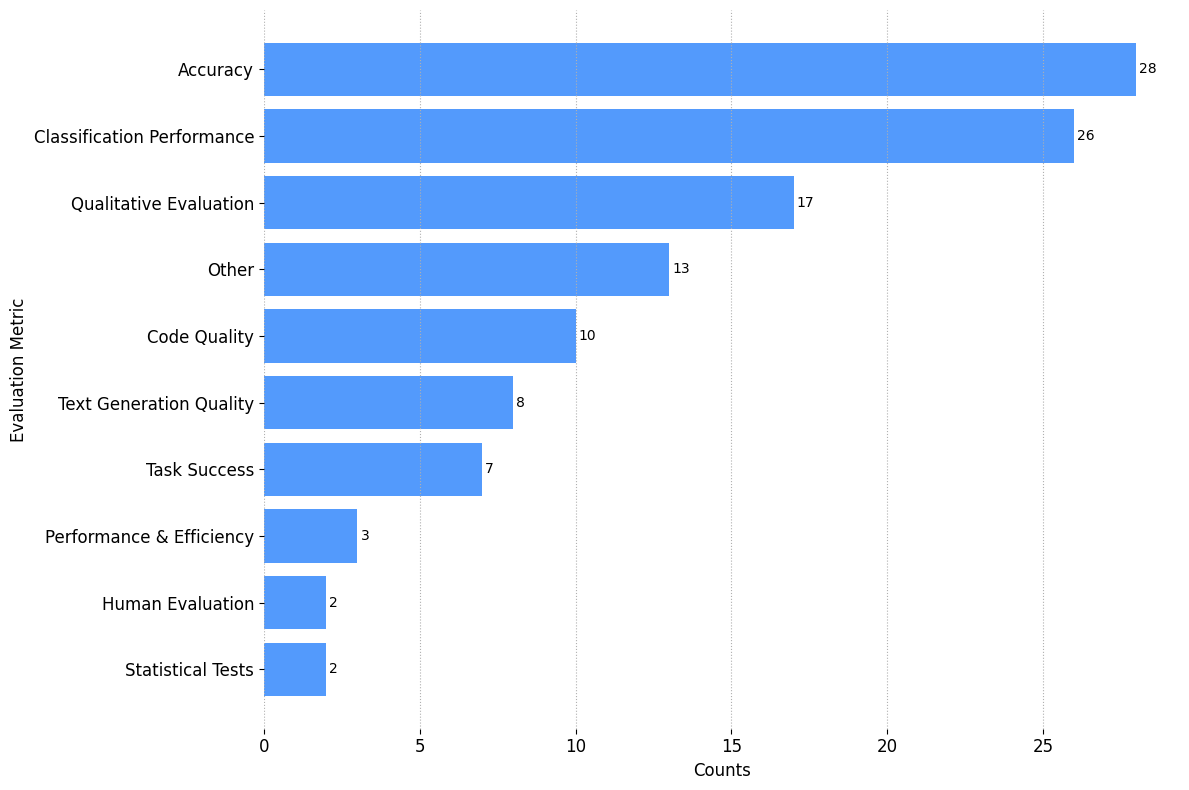}
    \caption{Frequency of Evaluation Metrics Used}
    \label{fig:rq2_eval_metric}
\end{figure}

Figure~\ref{fig:rq2_eval_metric} illustrates the evaluation metrics used in the analysis of LLMs. There were 37 individual evaluation metrics initially identified across research papers. To create a more interpretable visualization, similar metrics were systematically grouped into meaningful categories based on their evaluation purpose and domain.

Accuracy (used 24 times) and AUC/AUC-ROC (used 4 times) combined are the most frequently used metrics. Nirusanan et al. \cite{Nirusanan} utilized publicly available datasets from Kaggle to assess their model, measuring accuracy based on the percentage of correct responses across each dataset. Some studies \cite{jiang2024, schuster2024} conducted various queries to evaluate the model's performance, using accuracy to analyze the effectiveness of the models. In particular, Jiang et. al. \cite{jiang2024} conduct a case study and prompt the LLMs to generate SQL code of varying difficulty. They calculate the accuracy by classifying whether the answer provided by the LLM is acceptable.  

The Classification Performance category is consolidated into four traditional machine learning evaluation metrics: Precision (8), Recall (9), F1 score (8), and false positives/negatives (1). 

Bednár et. al \cite{bednar2023cognitive} cover problem and data understanding, and evaluation phases. They define recall as the number of correctly extracted concepts divided by the total number of correct concepts. Varma et al. \cite{varma2024reimagining} use precision, recall, and F1-scores to evaluate the performance of LLMs in Table Identification and Primary Key Identification (crucial steps in the Enterprise Data Management pipeline). 
Qualitative Evaluation \cite{ahn2024data, pido2023askyourdata, valverde-rebaza2024, zhao2024chat2data} was another commonly used metric and involved the verification of LLM responses by the authors or domain experts. 

The Text Generation Quality category combined four NLP-specific metrics (BLEU, ROUGE, BertScore, CodeBLEU) that evaluate the quality of generated text and code from a linguistic perspective. The Code Quality category unified four metrics specifically designed for code evaluation, including pass rate, \texttt{pass@1/Pass\_any@1}, execution accuracy, and correctness, focusing on functional code assessment.  The Task Success category merged three metrics that measure overall task completion, including success rate, exact match, and prompt completion rate. The Performance and Efficiency category combined three metrics related to computational resources and speed: computation time, efficiency/runtime, and cost. 
Finally, another category was created to group twelve diverse, low-frequency metrics that didn't fit into the main categories, including readability score, rationality, invalid/illegal rate, completeness, clarity, error propagation, similarity measurement, functional measurement, utility, Levenshtein distance, lift curve, and mean rank/best rank.  

Due to the relatively recent development of LLM evaluation methods, some studies have created their own assessment scales. For example, Huang et al. \cite{huang2024chatgpt} examined the efficiency of GPT-4.0 in epidemiology, performing a comparative analysis that assessed the model on factors such as consistency, analytical code efficiency, user-friendliness, and overall performance.

\subsection{RQ3: Large Language Models}

\begin{figure}[h]  
    \centering
    \includegraphics[width=0.48\textwidth]{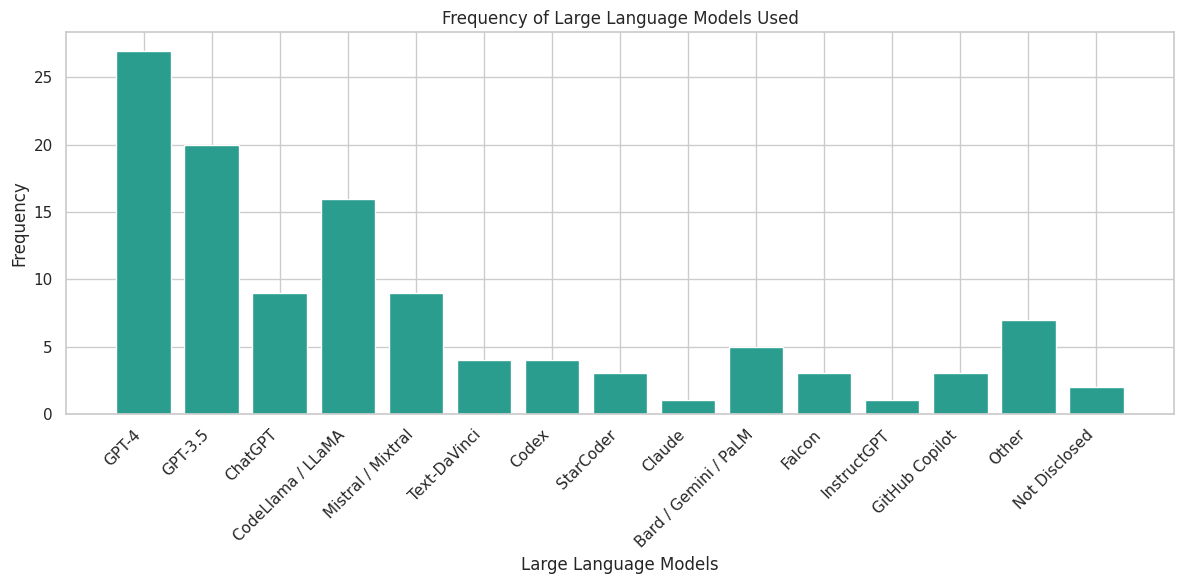}
    \caption{Large Language Models}
    \label{fig:rq3_llms}
\end{figure}

In analyzing the LLMs discussed in various papers, it is unsurprising that OpenAI's GPT series emerged as the leader. Many studies focused on GPT-4 and GPT-3.5, as shown in Figure \ref{fig:rq3_llms}. While most papers specified the exact version of the LLMs being examined, some simply referred to them as ChatGPT. Following the GPT series, Meta's Llama \cite{fang2024llms, gramacki2024, jiang2024, shinde2024, sudha2024}
series was a close second, with many papers comparing its performance to that of GPT. Additionally, a study analyzed the ChatGPT Code Interpreter \cite{ahn2024data}, a feature of GPT-4 specifically designed for programming tasks.

\subsection{RQ4: Application Domains}

\begin{figure}[h]  
    \centering
    \includegraphics[width=0.48\textwidth]{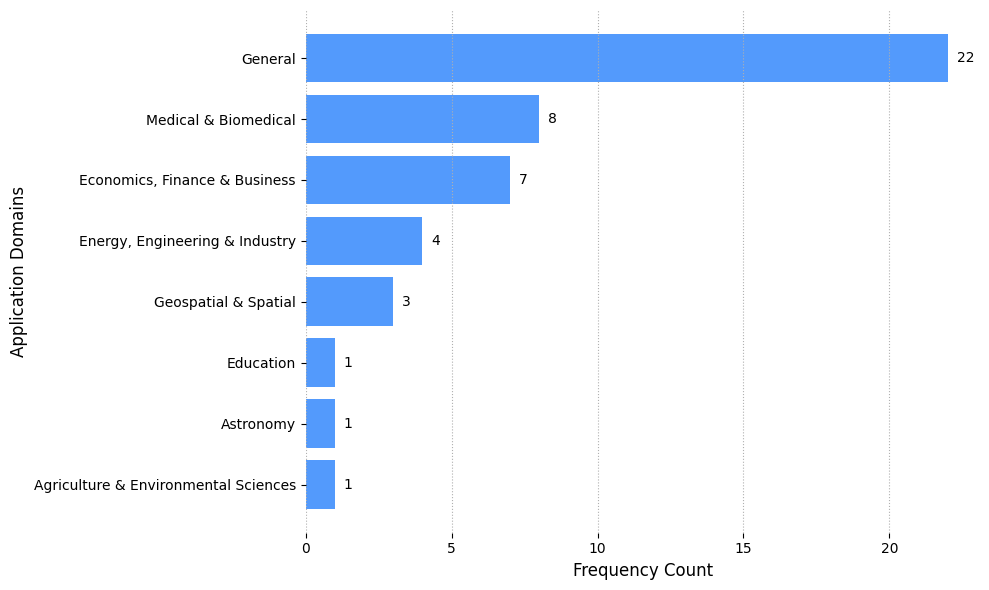}
    \caption{Application Domains}
    \label{fig:app_domain}
\end{figure}

The application domains of LLMs in data science span both general-purpose and domain-specific contexts. A significant portion of the literature addresses general data science challenges without targeting particular domains \cite{essabri2024, huang2024, jadoon2024contextmate, schuster2024, zhang2024benchmarking}. For instance, Zhang et al. \cite{zhang2024benchmarking} focus on fundamental data science operations, including data transformation, aggregation, cleaning, and manipulation, emphasizing broad applicability rather than domain-specific performance optimization.
Conversely, several studies examine LLM applications within specialized domains, such as finance \cite{korkanti2024enhancing} and healthcare \cite{ahn2024data}, demonstrating the technology's versatility across different fields, as showcased in Figure~\ref{fig:app_domain}. Clinical and medical applications \cite{ahn2024data, causio2024perspectives, sadeghi2023exploring, xie2024waitgpt} represent a notable area of focus, with research exploring how LLMs can streamline automated machine learning for clinical studies and broader AI implementation in healthcare settings. The study proposed in \cite{causio2024perspectives} exemplifies the growing interest in domain-specific applications within the healthcare sector.

Geospatial data science \cite{gramacki2024, jiang2024} has emerged as another prominent application area, with an increasing number of studies investigating LLM capabilities for spatial data analysis \cite{hojati2024large} and geographic information processing. Studies also focused on applying LLMs for data analysis in the Oil and Gas domain \cite{Seow2024KnowledgeAI}. This trend reflects the growing recognition of LLMs' potential to address complex, domain-specific analytical challenges beyond traditional general-purpose applications.
The distribution between general-purpose and domain-specific applications suggests that while LLMs demonstrate broad utility across data science tasks, researchers are increasingly exploring their specialized applications to leverage domain expertise and address field-specific analytical requirements.

\subsection{RQ5: Reported Positive and Negative Impacts}
\begin{figure}[h]  
    \centering
    \includegraphics[width=0.48\textwidth]{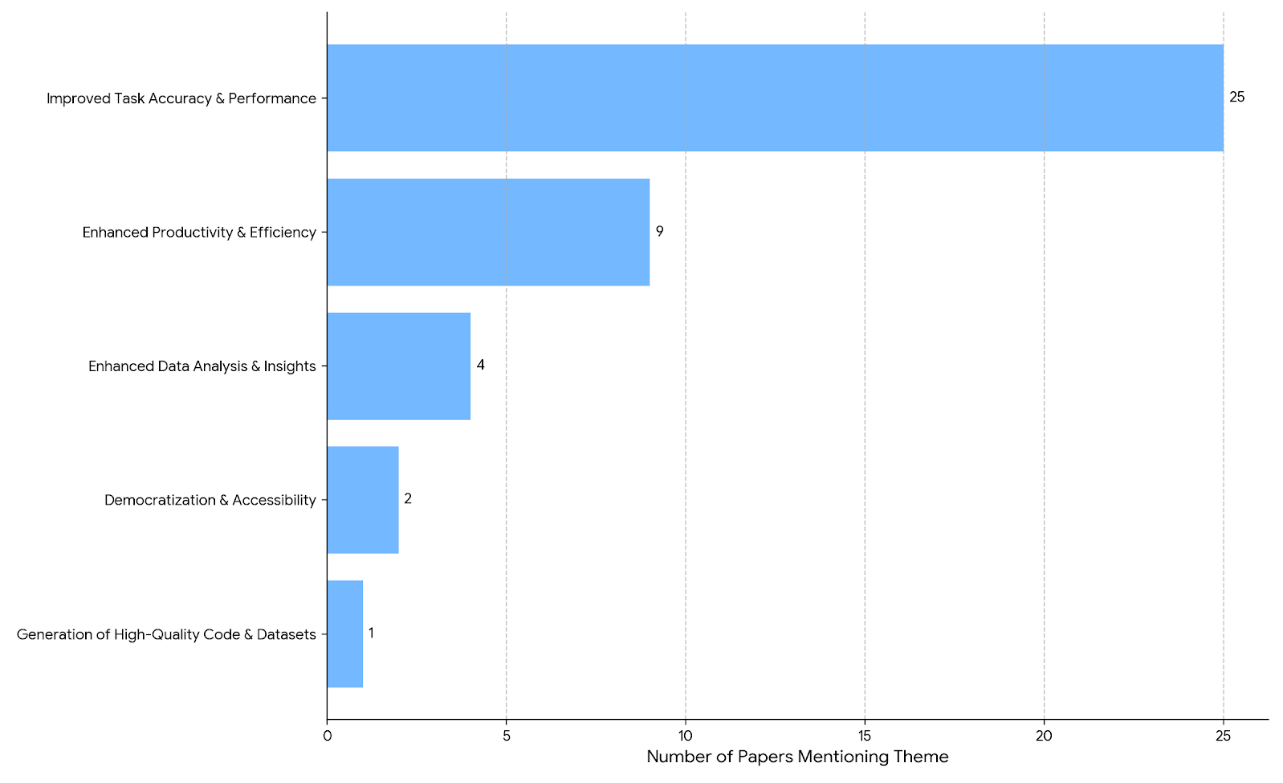}
    \caption{Frequency of Reported Positive Impact}
    \label{fig:frequency_positive_impact}
\end{figure}

The analysis of the literature was conducted on a corpus of 66 research papers. A keyword-based categorization was employed to identify recurring themes across three areas: positive impacts, negative impacts, and research gaps. As shown in Figure~\ref{fig:frequency_positive_impact}, this method successfully categorized the majority of the papers for each area, identifying positive impacts in 49 papers, negative impacts in 42 papers, and research gaps in 45 papers. The papers not captured in a specific category typically had no information provided for that field (e.g., no negative impacts were reported) or used highly specific language not matching the general keyword search. The following analysis, therefore, represents the distribution of themes found within these categorized subsets of the literature.

The collective research indicates that the most significant positive impact of LLMs in data science is the measurable improvement in task accuracy and performance. Across numerous studies, LLMs consistently outperform baseline models and, in some cases, even human experts. Beyond enhanced accuracy, these models are frequently reported to boost productivity and efficiency by automating complex tasks and reducing the time required for analysis. Furthermore, LLMs are democratizing the field by making sophisticated data science tools accessible to non-experts and are proficient at generating high-quality code, datasets, and nuanced insights that might otherwise be overlooked.
\begin{itemize}
\item Enhanced Productivity and Efficiency. Numerous studies report that LLMs significantly enhance the speed and efficiency of data science tasks. In a study by Jadoon et al. \cite{jadoon2024contextmate}, the ContextMate tool completed a task in 621.5 seconds, which was substantially faster than the baseline time of 798.9 seconds. Similarly, a system proposed by Pido et al. \cite{pido2023askyourdata} utilized only 14.85\% of the time required by the TPOT framework for classification tasks. The study by Varma et al. \cite{varma2024reimagining} found a 75\% reduction in the effort required for identifying source tables and columns, while another study by Zhao et al. \cite{zhao2025leva} noted that participants using an LLM-based tool completed tasks in less time than a control group.
\item Improved Task Accuracy and Performance. Several papers highlight the high accuracy of LLMs in performing various data science tasks. The study by Jadoon et al. \cite{jadoon2024contextmate} found that the tool's task accuracy was 93\%, a significant improvement over the baseline's 73\%. In another example, an ML model developed using ChatGPT outperformed a model created by experts in three out of four tasks \cite{arasteh2024langchain}. Furthermore, Guo et al. (2024) \cite{guo2024dsagentdatascience} discovered that their DS-Agent with GPT-4 achieved a 100\% success rate on the tested tasks, and Pido et al. (2023) \cite{pido2023askyourdata} reported that their DSBot achieved an accuracy greater than 95\% in six of their eighteen datasets.
\item Generation of High-Quality Code and Datasets. Several papers highlight the high accuracy of LLMs in performing various data science tasks. The study by Jadoon et al. \cite{jadoon2024contextmate} found that their tool's task accuracy was 93\%, a significant improvement over the baseline's 73\%. In another example, an ML model developed using ChatGPT outperformed a model created by experts in three out of four tasks \cite{arasteh2024langchain}. Furthermore, Guo et al. \cite{guo2024dsagentdatascience} found that their DS-Agent with GPT-4 achieved a 100\% success rate on the tasks they tested, and Pido et al. \cite{pido2023askyourdata} reported that their DSBot achieved an accuracy greater than 95\% in six of their eighteen datasets.
\end{itemize}

\begin{figure}[h]  
    \centering
    \includegraphics[width=0.48\textwidth]{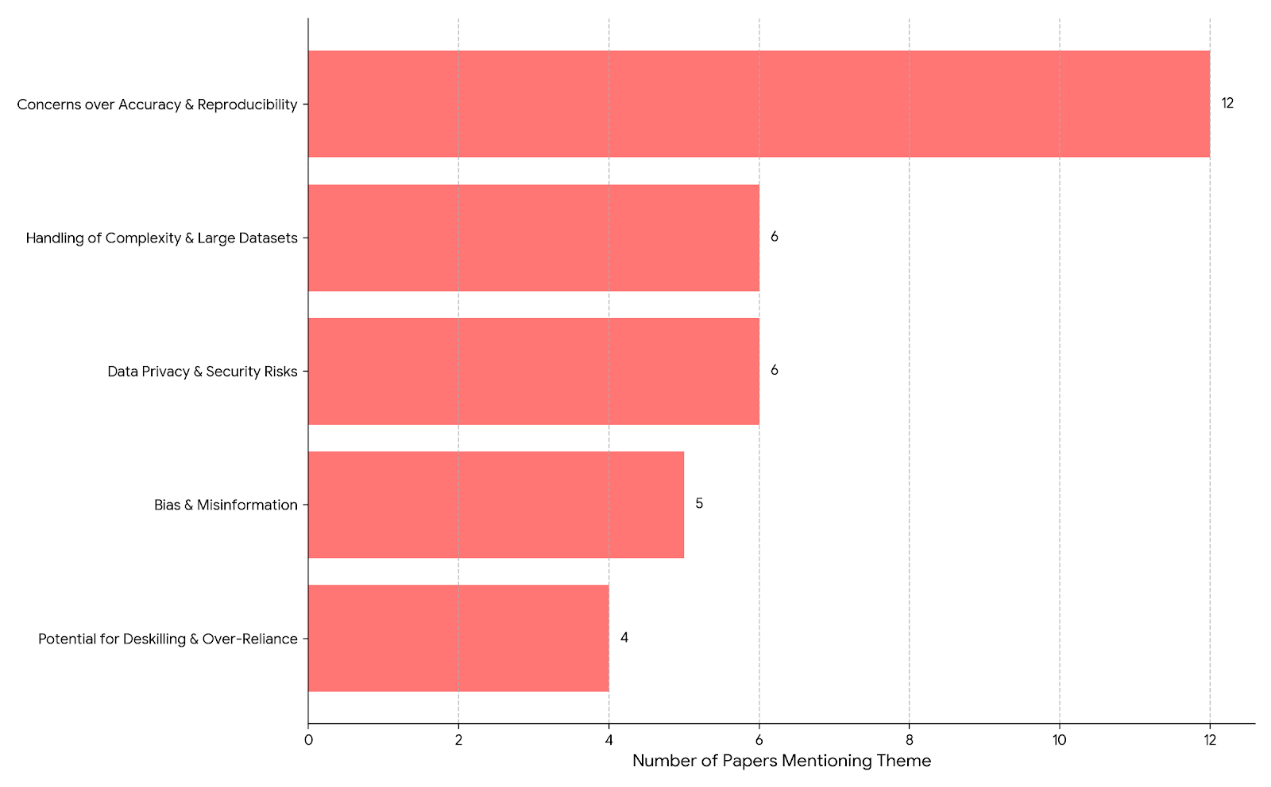}
    \caption{RQ5: Frequency of Reported Negative Impacts identified in 42 out of 62 papers}
    \label{fig:frequency_negative_impact}
\end{figure}

Despite their advantages, LLMs present several critical challenges. Figure \ref{fig:frequency_negative_impact} shows the most frequently cited negative impact in all papers, which are concerned with over-accuracy and reproducibility, as models can produce inconsistent, incorrect, or misleading results. Researchers also consistently highlight the difficulties that LLMs face when handling complex, open-ended tasks and large datasets, often due to inherent limitations such as token size. Other prominent concerns include significant data privacy and security risks, especially when using cloud-based APIs; the potential for user over-reliance, which could hinder the development of critical thinking skills; and the risk of generating biased content.
\begin{itemize}
\item Concerns over Accuracy and Reproducibility. A primary concern is the reliability of LLM outputs. In an evaluation by Patel and Shooshtarian \cite{patel2024}, some visualizations, such as a confusion matrix, were plotted without any results. Another study conducted by Mishra et al. \cite{mishra2024dataagent} found that LLMs sometimes generated code with variables that did not exist in the dataset. The non-deterministic nature of LLMs is also a key issue; Lanfermann et al. \cite{lanfermann2024llm} and Maddigan and Susnjak \cite{maddigan2023chat2vis} both found that slightly different or even identical prompts can lead to varying results, which undermines reproducibility.
\item Data Privacy and Security Risks. The use of LLMs, particularly cloud-based services, raises significant concerns regarding data privacy and security. A study by Huang et al. \cite{huang2024} explicitly stated that data privacy is a concern when using ChatGPT for epidemiological data analysis. This is echoed in other studies by Ghazzai et al. \cite{ghazzai2024harnessing} that highlight privacy issues related to sending data to external servers. These concerns are particularly pronounced for sensitive information, with one study \cite{ahn2024data} noting that the use of API-based chatbots for biomedical applications can lead to privacy issues.
\item Handling of Complexity and Large Datasets. LLMs are reported to have difficulty with complex, open-ended data science tasks and large datasets. The study by Zhang et al. \cite{zhang2024elfgym} noted that their paper did not include high-level, open-ended, and vague data science tasks. Similarly, \cite{Nirusanan} found that their model was not skilled at handling highly complex financial tasks. Other researchers \cite{jiang2024, patel2024} have observed that LLMs struggle with large datasets and that their performance tends to decline as the complexity of the questions increases.
\end{itemize}

\begin{figure}[h]  
    \centering
    \includegraphics[width=0.48\textwidth]{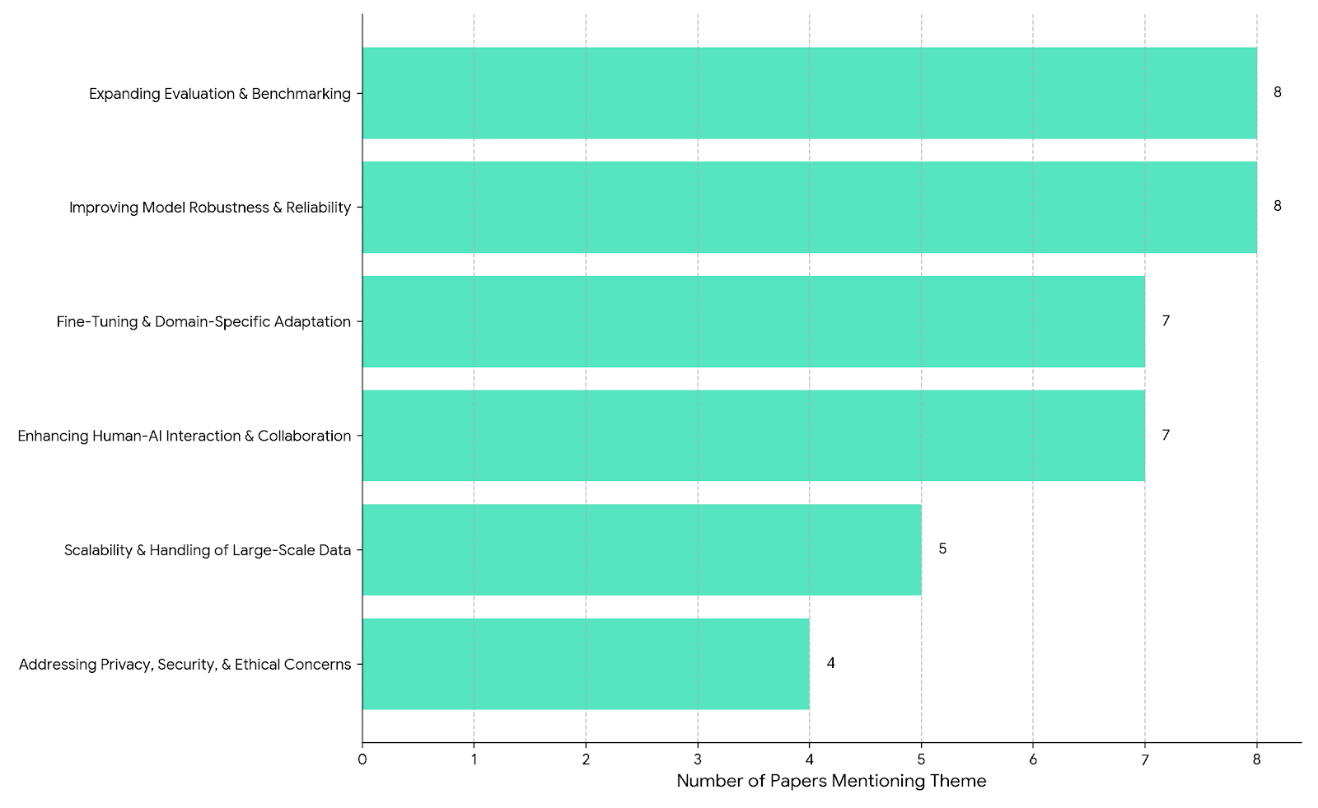}
    \caption{RQ5: Frequency of Reported Research Gaps identified in 45 out of 62 papers}
    \label{fig:frequency_research_gaps}
\end{figure}

\subsection{RQ6: Reported Research Gaps}

The analysis of the literature reveals a clear consensus on directions for future work. The most pressing research gaps are the need to expand evaluation and benchmarking to better assess LLM capabilities on complex, real-world tasks and the critical need to improve model robustness and reliability (see Figure \ref{fig:frequency_research_gaps}. Researchers frequently call for enhancing human-AI interaction to create more intuitive and user-friendly interfaces. Other consistently mentioned gaps include addressing challenges in scalability to handle large-scale data, developing more effective methods for fine-tuning and domain-specific adaptation, and addressing the persistent ethical concerns surrounding privacy, security, and algorithmic transparency.
\begin{itemize}
\item Improving Model Robustness and Reliability. A significant number of papers call for future work to focus on enhancing the robustness and reliability of LLMs. This includes improving models for data preparation and analysis tasks \cite{nasseri2023LLM} and addressing issues such as high response latency and token limits, which negatively affect the user experience \cite{jadoon2024contextmate}. Further proposed research includes the critical evaluation of results and prompts \cite{foltyn2024gpt4}, as well as the need for additional training and adaptation for LLMs to become viable tools for specialized tasks, such as cleaning semi-structured errors \cite{mondal2024cleaning}.
\item Expanding Evaluation and Benchmarking. Many studies suggest the need for a more comprehensive evaluation of LLMs. One study noted that even the best-performing models have accuracy levels that are far from desirable, indicating a need for better benchmarks \cite{huang2024}. Future work proposals include expanding benchmarks for specialized areas like geospatial code generation \cite{gramacki2024}, creating larger datasets for evaluation \cite{chen2025viseval}, and assessing LLMs on more complex, real-world data science tasks with expanded evaluation metrics \cite{nascimento2025llm}.
\item Enhancing Human-AI Interaction and Collaboration. Future research should explore ways to better support users interacting with LLMs. Proposals include designing AI interfaces that enable adaptive decomposition responsive to the user and task \cite{Kazemitabaar_2024} and examining how AI assistance can be customized to support novice data analysts \cite{gu2024respond}. Additional research gaps involve enhancing the conversational capabilities of chatbots to facilitate a more natural interaction \cite{pido2023askyourdata} and improving visual interfaces to democratize data consumption with verifiable AI \cite{xie2024waitgpt}.
\item Addressing Privacy, Security, and Ethical Concerns. There is a clear need for further research into the ethical implications of using LLMs in data science. Future work should address concerns related to privacy, security, potential bias, and intellectual property issues \cite{mohammed2023metadata}. Other proposed research directions include exploring privacy preservation and algorithmic transparency \cite{varma2024reimagining} and giving careful consideration to the ethical implications and potential biases when integrating LLMs into new domains \cite{sadeghi2023exploring}.
\end{itemize}

\subsection{Other Discussions}

This mapping study revealed a rapidly evolving field where LLMs are being systematically integrated across all phases of the data science pipeline, with the predominance of empirical studies (87.7\%) demonstrating the field's practical orientation and commitment to real-world validation. The consistent pattern of framework development followed by evaluation, exemplified by several studies (e.g., DSAgent \cite{guo2024dsagentdatascience}, LLM4V \cite{beasley2024pipeline}, ARGUMENT2CODE \cite{zhao2024new}) highlights the methodological rigor of current research. However, the findings reveal a nuanced landscape where LLMs demonstrate clear utility in specific tasks while facing significant limitations that prevent autonomous operation. The comparative analysis by Bednár et al. \cite{bednar2023cognitive} illustrates this complexity, showing that while human data scientists excel in identifying domain-specific technical KPIs, GPT-4 demonstrates superior performance in task definition and problem articulation.

The evaluation landscape reflects both the maturity and limitations of current LLM applications in data science. While accuracy dominates as the primary evaluation metric, the emergence of evaluation frameworks that assess many factors (e.g., consistency, code efficiency, user-friendliness) indicates a recognition that traditional metrics may be insufficient for capturing the full spectrum of LLM performance. The OpenAI's GPT series, particularly GPT-4 and GPT-3.5, outperform other LLMs \cite{essabri2024, ghazzai2024harnessing, zhang2024benchmarking}. Despite these advances, fundamental challenges persist, including concerns about accuracy and reproducibility where identical prompts can produce varying results \cite{nejjar2025llms}, significant data privacy and security risks \cite{mohammed2023metadata}, and difficulties handling complex, open-ended tasks due to inherent limitations such as token size constraints \cite{shinde2024}.

%% file: sections/5-implications.tex
\section{Implications}

This section discusses the implications of the study for the research community. 

\subsection{Implications for Researchers}
This study presents a comprehensive examination of how LLMs are applied across the data science lifecycle. By organizing findings according to the distinct phases of the data science process and highlighting relevant subtopics, the study provides a structured foundation for future academic inquiry. The automation of data analysis tasks using LLMs is the most explored area, addressed in over 40\% of the reviewed studies. In contrast, critical downstream stages, such as deployment and model maintenance, remain significantly under-investigated, with fewer than 1\% of studies focusing on them. This disparity reveals a clear opportunity for further research.

Although LLMs demonstrate potential in improving accuracy, increasing efficiency, and enabling automation, their non-deterministic behavior continues to pose a challenge. Several researchers have attempted to mitigate this issue through strategies such as structured prompting, response sampling, and aggregation techniques. However, there remains a pressing need for more robust methods to ensure consistency and reliability in LLM outputs. Future work could also investigate mechanisms for integrating LLMs more reliably into data science pipelines, especially in high-stakes or regulated environments.

\subsection{Implications for practitioners}
While the primary aim of this study is to inform and guide academic research, it also presents key insights for practitioners seeking to adopt LLM-based tools in real-world data science workflows. Our study identifies several innovative frameworks, such as HuggingGPT \cite{shen2023hugginggpt}, WaitGPT \cite{xie2024waitgpt}, Elf-gym \cite{zhang2024elfgym}, MLCopilot \cite{zhang2024mlcopilot}, and CAAFE \cite{hollmann2023caafe}, that leverage LLMs to streamline various data science tasks. These frameworks can serve as starting points for practitioners interested in exploring the integration of LLMs in their processes.

Additionally, several benchmarks have been introduced to evaluate LLMs at specific stages of the data science lifecycle. For instance, SpreadNaLa \cite{schuster2024} evaluates spreadsheet formula generation, InfiAgent-DABench \cite{hu2024infiagent} assesses capabilities in data analysis, and VisEval \cite{chen2025viseval} focuses on visualization tasks. These resources provide practitioners with valuable tools for evaluating the effectiveness of LLMs in targeted applications.

It is worth noting that while some frameworks have been implemented and made publicly available \cite{guo2024dsagentdatascience}, others remain in the prototype stage \cite{whitehead2024generative}, offering practitioners a pathway to contribute by deploying, extending, or adapting these tools for production environments. However, practical constraints, such as token length limitations reported in studies \cite{ghazzai2024harnessing, jadoon2024contextmate, arasteh2024langchain}, must be carefully considered, particularly in use cases involving large datasets or complex workflows.

Overall, this study highlights the increasing applicability of LLMs in data science, while also identifying critical areas where further development, validation, and real-world testing are necessary to ensure their effective and reliable use.

%% file: sections/6-limitations-and-treats.tex
\section{Limitations and Threats to Validity}

\subsection{Internal Threats}
\subsubsection{Bias in Data Synthesis}
A key challenge during the data extraction and synthesis process was the lack of complete or consistent information across the reviewed studies. For instance, some studies referenced the use of ChatGPT without specifying the version, which can significantly impact interpretation, given the rapid evolution of LLM capabilities. The absence of this detail limits the precision of comparative analysis across studies.

Additionally, several studies did not disclose the datasets used for evaluating their proposed frameworks or experiments. In some cases, authors cited privacy concerns as the reason for withholding dataset details \cite{foltyn2024gpt4, zhang2024}, further complicating efforts to validate or replicate their findings. These omissions introduce potential bias in the synthesis process and limit the reproducibility and generalization of our analysis.

To mitigate such issues, we employed a structured template for data extraction and engaged in iterative discussions among reviewers to reach consensus wherever inferences were required. However, we recognize that the quality and completeness of the underlying source material ultimately constrain the quality of synthesis.

\subsection{External Threats}
\subsubsection{Search String Limitation}

We developed a comprehensive search string to capture as many relevant studies as possible; however, the broad and evolving nature of the field—particularly in the application of large language models (LLMs) across the data science lifecycle—means that some important studies were likely missed. This challenge is compounded by the fact that data science is a relatively new and rapidly developing discipline without a universally accepted definition of its lifecycle. Researchers often use different frameworks and terminology to describe the stages and tasks involved, resulting in the same stage being referred to by multiple names and activities within each stage being described inconsistently.

To address this challenge, we incorporated a range of synonymous terms into our search strategy to capture the diversity of language used in the field. For instance, for visualization tasks, we included terms such as “data visualization,” “chart generation,” and “plot creation”; for programming assistance, we used “code generation,” “script generation,” and “automated coding”; for data preparation activities, we added “data preparation,” “data cleaning,” “data wrangling,” “data preprocessing,” and “ETL”; and for analytical processes, we used “data analysis,” among others. 

Despite these efforts, the inconsistent naming conventions and the interdisciplinary nature of the topic suggest that some papers may have employed highly specific terminology not covered by the general or semi-specific terms in our query. To mitigate this limitation, we configured the search to identify matches in both titles and abstracts. Nevertheless, we acknowledge that some relevant studies may still have been missed, particularly given the emerging nature of the field and the varied ways researchers describe LLM applications across different stages of the data science lifecycle.

%% file: sections/7-conclusion.tex
\section{Conclusion}

Incorporating LLMs into data science workflows marks a notable shift in how analytical work is done, offering real-world opportunities to boost productivity and make data analysis more accessible to non-experts. Current research suggests that the key to success lies in positioning LLMs as supportive tools that work in conjunction with human analysts, rather than as standalone solutions. The field has made significant progress in developing evaluation methods, validation approaches, and benchmark datasets, solving both general-purpose and domain-specific problems, which provides a solid foundation to build upon. However, significant hurdles remain around ensuring accuracy for complex tasks, maintaining reproducible results, and protecting sensitive data. These are issues that need to be resolved before organizations can confidently utilize LLMs for mission-critical analysis. 